\documentclass[aps,prl,twocolumn,groupedaddress]{revtex4}  
\usepackage{graphicx}  
\usepackage{dcolumn}   
\usepackage{bm}        
\usepackage{amssymb}   
\usepackage{natbib}
\usepackage{epstopdf}
\usepackage{amsmath}
\usepackage{caption}
\usepackage{subcaption}
\usepackage{float}
\restylefloat{table}

\global\long\def\abs#1{\left\vert #1\right\vert }

\global\long\def\ket#1{\left\vert #1\right\rangle }

\global\long\def\co2{\text{CO}_{2}}

\begin{document}

\title{Measurement of the spin-dipolar part of the tensor polarizability of $^{87}$Rb }
\author{Yehonatan Dallal and Roee Ozeri\\
\small \emph{Department of Physics of Complex Systems, Weizmann Institute of Science, Rehovot 7610001, Israel} }
\date{July 31,2014}

\begin{abstract}
We report on the measurement of the contribution of the magnetic-dipole hyperfine interaction to the tensor polarizaility of the electronic ground-state in $^{87}$Rb.
This contribution was isolated by measuring the differential shift of the clock transition frequency in $^{87}$Rb atoms that were optically trapped in the focus of an intense CO$_2$ laser beam. By comparing to previous tensor polarizability measurements in  $^{87}$Rb, the contribution of the interaction with the nuclear electric-quadrupole moment was isolated as well.
Our measurement will enable better estimation of black-body shifts in Rb atomic clocks. The methods reported here are applicable for future spectroscopic studies of atoms and molecules under strong quasi-static fields.
\end{abstract}

\maketitle

Atomic systems are a good experimental platform for the test of quantum many-body theories with high precision. Atomic structure calculations for heavy atoms, with a large number of electrons, require sophisticated approximations and numerical methods. The predictions of such calculations can be readily tested using spectroscopy experiments. One such prediction entails atomic polarizabilities. 

Due to their symmetry under parity, atoms do not have permanent electric-dipole moments. When placed in a static electric field, however, their electronic wavefunction is polarized and a dipole moment which is proportional to the applied field is induced. The atomic polarizability is the proportionality tensor between the induced dipole moment and the field. This tensor can be further reduced to scalar and rank-two tensor parts. The latter, also known as  the tensor polarizability, is determined by the hyperfine interaction of the electron with different magnetic and electric moments of the nucleus. The ab-initio calculation of the different polarizabilites is a difficult task and requires the use of advanced quantum many-body methods \cite{DzubaFlambaumDerevianko_PRA2010}. The measurement of the tensor atomic polarizability requires precision spectroscopy under strong applied electric fields.

The precision of spectroscopic experiments benefits from long interrogation times. Optical dipole traps with their long storage times were therefore considered as a promising platform to this end. However, it was soon realized that the perturbation of atomic levels by the trapping optical fields introduces systematic line shifts and broadenings, and therefore compromises the precision of spectroscopy. Here, we rather take advantage of the large Stark shifts of atoms that are trapped in the focus of an intense CO$_2$ laser field in order to measure the tensor differential polarizability of the clock transition in $^{87}$Rb. The slowly varying field of the CO$_{2}$ laser, as compared with the atomic resonance frequencies in Rb, allows for the measurement of the static polarizability to a good approximation. The interaction of atoms with laser light has been previously used to investigate atomic structure \cite{Rosenband2006, Sherman2008, Porsev2008, Holmgren2012}

This tensor shift of the clock transition depends only on the contribution of the spin-dipolar hyperfine interaction to the polarizability and therefore allows for its determination \cite{SandarsAngel_RSPA1968}. A comparison with a previous measurement of another tensor polarizability in $^{87}$Rb enables the isolation of the nuclear electric-quadrupole contribution as well. Both results confirm recent atomic structure calculations \cite{DzubaFlambaumDerevianko_PRA2010}.

For atomic clocks the shift due to the stochastic field of black body radiation (BBR) is of concern. Uncertainty in the BBR shift is one of the limiting factors in the accuracy of Rb based atomic clocks \cite{GuenaClarionRosenbusch_IEEE2010,Safronova_PRA2010}.
For an isotropic field such as BBR, the tensor shift averages to zero so that only the scalar shift is relevant.
Nevertheless, since when measuring the shift under a static field to determine the scalar polarizability, the two effects occur together, the value of the tensor polarizability must be precisely known to correctly account for it \cite{Mowat_PRA1972,SimonClarion_PRA1998}.

An alkali atom in its electronic ground state, when placed in a static electric field, experiences a shift, $h\delta$, of its energy levels due to the Stark effect \cite{UlzegaThesis, UlzegaWeis2006},
\begin{eqnarray}
&& h\delta_{F=I+\frac{1}{2}}=-\frac{1}{2}|\textbf{E}|^2(\alpha+ \alpha_{10}+\\ \nonumber
&& (\alpha_{12}+\alpha_{02})\frac{3M^{2}-F(F+1)}{I(2I+1)}\frac{3\cos^{2}\theta-1}{2}),\\
&& h\delta_{F=I-\frac{1}{2}}=-\frac{1}{2}|\textbf{E}|^2(\alpha-\frac{I+1}{I}\alpha_{10}+\\ \nonumber
&& (-\alpha_{12}+\frac{2I+3}{2I-1}\alpha_{02})\frac{3M^{2}-F(F+1)}{I(2I+1)}\frac{3\cos^{2}\theta-1}{2}).
\end{eqnarray}

Here $F$ is the total angular momentum of the state, $M$ is its projection on the quantization axis, $I$ is the nuclear angular momentum, $\alpha$ is the polarizability in the absence of hyperfine interactions and $\alpha_{10}$ is the scalar differential polarizability. The contribution of the spin-dipolar and the electric quadrupole hyperfine interactions to the tensor polarizabilty are $\alpha_{12}$ and $\alpha_{02}$ correspondingly. The applied electric field magnitude is $|\textbf{E}|$ and $\theta$ is the angle between this field and the quantization axis.

 All previous measurements of the tensor polarizability in alkali atoms examined the dependence of the Stark shift on $M$, by detecting the shift of the flop-in transition under an applied electric field \cite{GouldWeisskopf_PR1969,OpelklausWeis2003}. This shift is proportional to $\alpha_{12}+ \alpha_{02}$. Here, we measured the dependence of the differential Stark shift of the clock transition; i.e. the transition between the $M=0$ states of the two hyperfine manifolds; on $\theta$. In the case of $^{87}$Rb, $I=3/2$, $F=2$, $1$ and the differential shift of the clock transition, $|F=1,M=0\rangle \rightarrow |F=2, M=0\rangle$, is simply,
\begin{equation}
h\delta_{clock}(\theta) = -\frac{4}{3}\alpha_{10}|\textbf{E}|^2 + \frac{1}{3}\alpha_{12}(3\cos^{2}\theta-1)|\textbf{E}|^2.
\label{eq:clock_shift_EB_ang}
\end{equation}
We measured the difference in the Stark shift between two perpendicular orientations of the magnetic quantization field with respect to the electric field direction; i.e. laser polarization,
\begin{equation}
h\delta_T=h\delta_{clock}(\theta=0) - h\delta_{clock}(\theta=\pi/2)=\alpha_{12}|\textbf{E}|^2.
\label{eq:clock tensor_shift}
\end{equation}
Thus, our measured shift depends solely on the spin-dipolar contribution to the tensor polarizability. Note that $\delta_T$ is typically two orders of magnitude smaller than $\delta_{clock}$.


\begin{figure}
\centering
\includegraphics[width=0.85\linewidth]{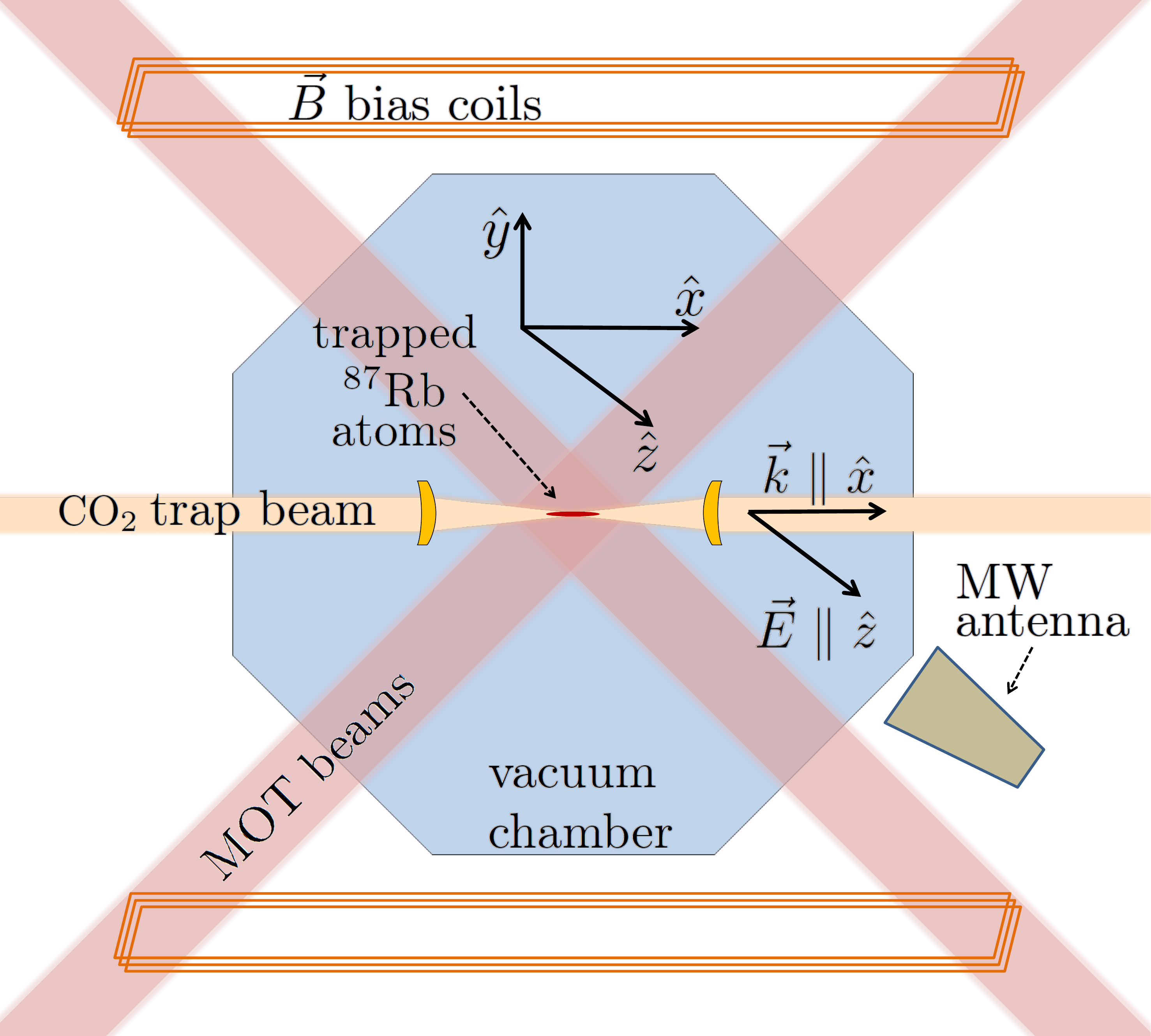}
\caption{Experimental system diagram.
The trap beam is propagating along the $\hat{x}$ axis and is linearly polarized along the $\hat{z}$ axis.
During the experiment the magnetic field can be applied along each of the major axes using three perpendicular pairs of magnetic bias coils.
Two pairs of magnetic coils were omitted from the diagram.}
\label{fig:figure1}
\end{figure}

In our experiment we used microwave Ramsey spectroscopy of atoms trapped in a quasi-electrostatic $\co2$ trap to measure the differential Stark shifts under a strong electric field.
Here the $\co2$ laser serves both as a trapping potential and a means to apply strong electric fields in vacuum.
While the electric field of a $\co2$ laser operating at a wavelength of 10.6$\mu$m is not static, it can be treated as quasi-static due to the large detuning from the closest dipole transition in Rb, at 0.795$\mu$m \cite{TakekoshiKnize_OPtCom1995}.
The relative difference between the static polarizability and that at 10.6$\mu$m is on order of 1\% \cite{Safronova_PRA2010}.

In our setup the trapping beam wave vector $\vec{k}$ is along the $\hat{x}$ direction, and is linearly polarized along the $\hat{z}$ direction ($\textbf{E}\parallel\hat{z}$), whereas $\hat{y}$ is perpendicular to both, see Fig.\ref{fig:figure1}.
With the axes defined, $\theta=\pi/2$ for $\textbf{B}\parallel\hat{x},\hat{y}$ and $\theta=0$ for $\textbf{B}\parallel\hat{z}$.
By repeating the measurement for various trapping beam intensities we obtain $\delta_{clock}$ and $\delta_{T}$ as a function of laser power, or equivalently $\abs{\textbf{E}}^2$.

We now turn to describe the experimental procedure.
Atoms were collected and cooled by a six-beam magneto optical trap and were then loaded into a trap formed by a focused $\co2$ beam.
The beam waist at the focus was $30\mu$m and its power was varied between 0.75W and 4.5W, resulting in an electric field on the order of 1MV/m.
The power of the laser was actively stabilized to a 1\% level.
After loading, all cooling lasers were extinguished and a self-evaporation phase followed, allowing the $1-4\times10^4$ atoms to cool to a temperature of 10-30$\mu$K, depending on the trap depth.
After one second of self-cooling we applied a magnetic field along the $\hat{x}$ direction.
Using a combination of microwave and laser pulses we optically-pumped the atoms to the $\ket{F=1,m=0}$ state.
We then adiabatically rotated the magnetic field to a chosen direction by adjustments to the bias coils currents.
The direction of the applied magnetic field was cycled from one data point to the next between $\hat{x}$, $\hat{z}$ and $\hat{y}$. An additional delay of two seconds was introduced to allow for eddy currents and control transients to relax before spectroscopy begins.

\begin{figure}
\centering
\includegraphics[width=0.95\linewidth]{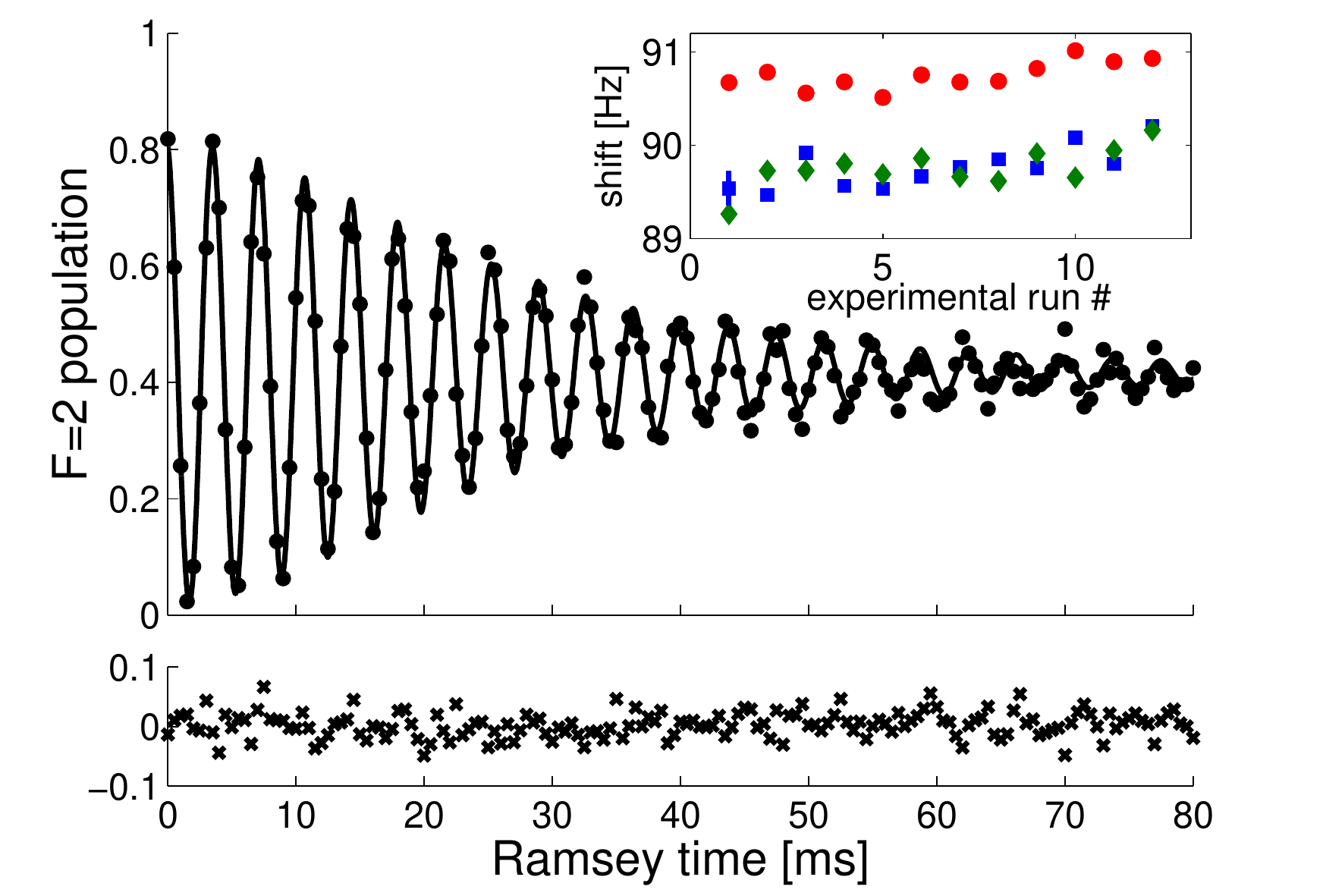}
\caption{Ramsey signal, $R(t)$, recorded for a trap laser power of 1.5W and microwave clock tuned to 350Hz bellow the clock transition frequency.
The measured data is shown by the black filled-circles. A fit to Eq.\ref{eq:fit_func} is shown by the black solid line. From the fit we obtain the decay constant $K$ and the detuning $\delta$.
The residuals are plotted below.
\textit{Inset}: $|\delta_{clock}|$ as extracted from 12 successive runs at the same trap power. The three sets correspond to the directions of the applied magnetic field: blue squares $\mathbf{B}\parallel\hat{x}$, green diamonds $\mathbf{B}\parallel\hat{y}$ and red circles $\mathbf{B}\parallel\hat{z}\parallel\textbf{E}$. The vertical line on the left-most blue square represents the typical 95\% confidence interval for a single point.
A tensor shift of approximately 1\,Hz is seen as the difference between the $\hat{x}$,$\hat{y}$ and $\hat{z}$ datasets. The drift in the measured shift is due to a drift in trapping laser power, and is smaller than 1\%.}
\label{fig:figure2}
\end{figure}

The different Stark shifts were measured using Ramsey spectroscopy. The time between the two $\pi/2$ pulses, $t$, was scanned and the population in $F=2$ following the second $\pi/2$ pulse was measured by absorption imaging.
The total number of atoms in the trap was subsequently measured after 1 ms.
The fraction of atoms in the $F=2$ level after the Ramsey pulse, $R(t)$, is shown in Fig.\ref{fig:figure2} in black filled-circles. A maximum-likelihood fit is performed with respect to a theoretical curve for non-interacting atoms in a harmonic potential, shown by the black solid line \cite{KuhrMeschede_PRA2005},
\begin{multline}
R(t)=C+\\ A\left[1+\left(\frac{t}{K}\right)^{2}\right]^{-\frac{3}{2}}\cos\left[2\pi\delta t-3\arctan\left(2\pi\frac{t}{K}\right)+\varphi\right]+S\cdot t\label{eq:fit_func}.
\end{multline}
Here the bias, $C$, and the amplitude, $A$, should equal $\frac{1}{2}$ for a perfect initialization. A phenomenological linear slope, $S$, is observed in the data and can be attributed to a slow population relaxation process.
Had $S\cdot t$ not been included in the model, the effect would have been observed as a slope in the residuals, without a statistically significant effect on the estimated shift.
The signal $R(t)$ decays with a typical time-scale, $K=\frac{2h}{\eta k_{B}T}$, where $\eta$ is the ratio of $h\delta_{clock}$ to the trap depth, $\frac{1}{2}\alpha|\textbf{E}|^2$.
The detuning of the clock transition relative to the local oscillator frequency is $\delta=\delta_{LO}-(\delta_{clock}+\delta_{n}+\delta_{B})$. Here, $\delta_{LO}$ is the microwave source detuning from the free $^{87}$Rb clock transition frequency, $\delta_{B}=\abs{\textbf{B}}^2\times57.515\text{kHz/mT}^{2}$ is the second order Zeeman shift of the clock transition at a magnetic field B and $\delta_{n}$ is the density dependent collisional shift.
The detuning, $\delta$, as well as the decay constant, $K$, were estimated by the fit.
The differential Stark shift, $\delta_{clock}$, was evaluated after subtracting $\delta_B$ and $\delta_n$.

The measurement was repeated 12 to 16 times for each of the applied magnetic field directions at different trapping laser powers, covering half a decade of Stark shifts.
The inset of Fig.\ref{fig:figure2} shows the results of one such measurement at a fixed trap power.
The difference in shifts between the measurements performed with $B\parallel\hat{z}$, and those performed with $B\parallel\hat{x},\hat{y}$ equals $\delta_T$.

To relate the measured shifts to the differential polarizabilities one needs to know the electric field magnitude.
Here, by evaluating the ratio,
\begin{equation}
m \equiv \frac{\delta_T}{\delta_{clock}(\theta=0)} = -\frac{3\alpha_{12}}{4\alpha_{10}-2\alpha_{12}},\label{eq:measurable_s_t_ratio}
\end{equation}
we use the known $\alpha_{10}$ \cite{Safronova_PRA2010, Mowat_PRA1972} and the atoms themselves as probes of the laser electric field.

\begin{figure*}
\centering
\begin{subfigure}{.5\textwidth}
  \centering
  \includegraphics[width=0.8\linewidth]{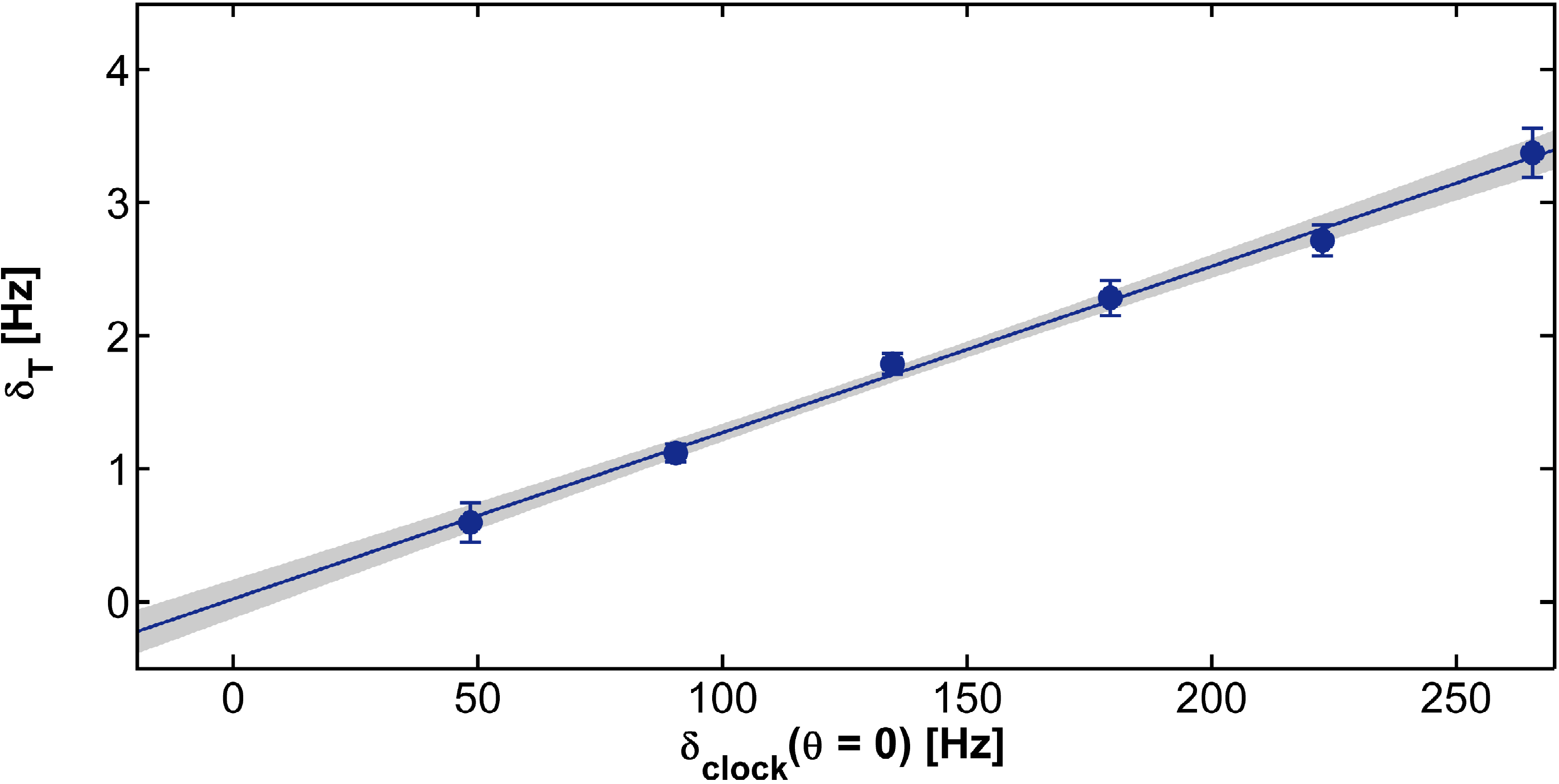}
  \caption{}
  \label{fig:sub1}
\end{subfigure}%
\begin{subfigure}{.5\textwidth}
  \centering
  \includegraphics[width=0.8\linewidth]{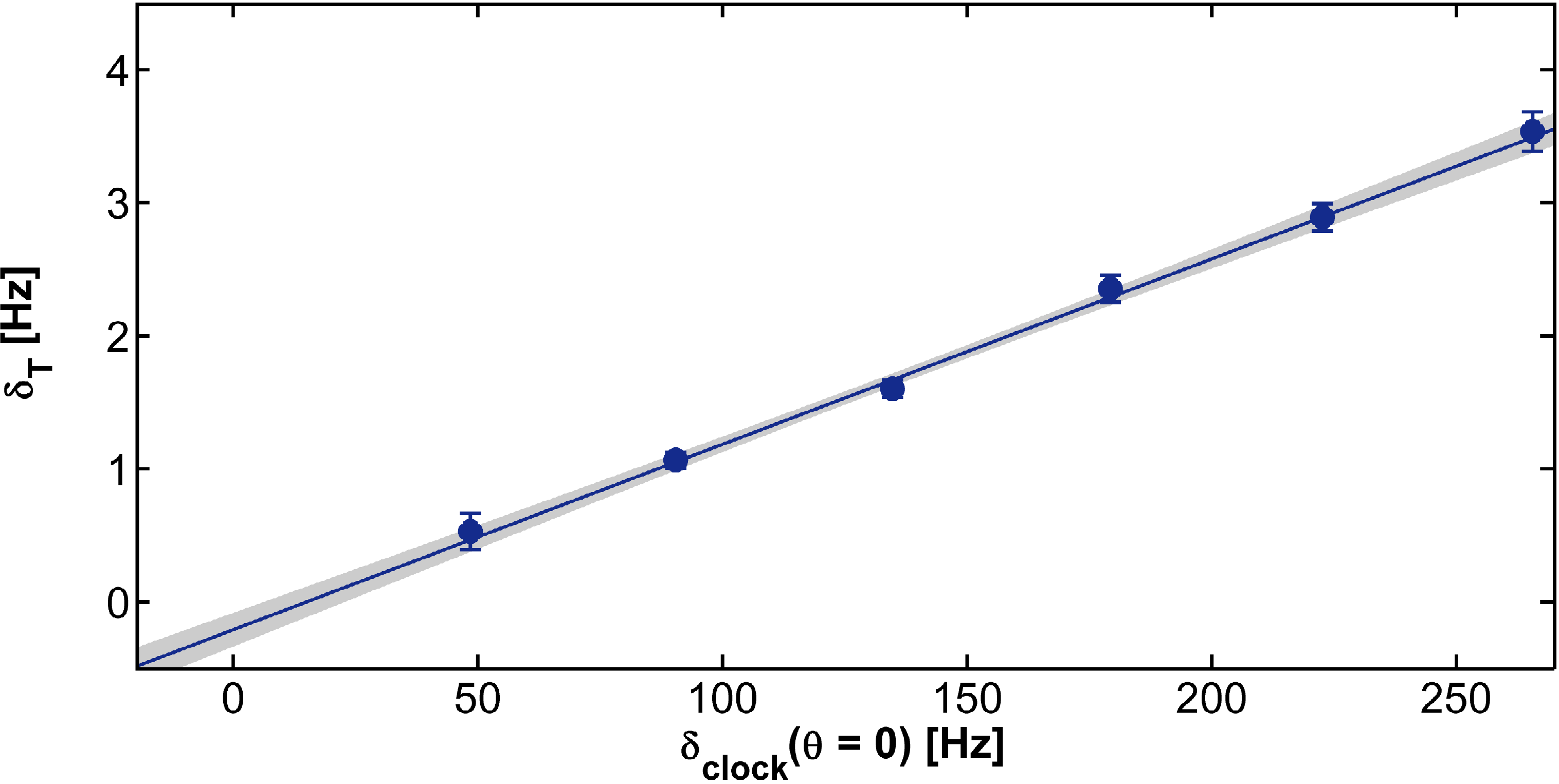}
  \caption{}
  \label{fig:sub2}
\end{subfigure}
\caption{Tensor shift $\delta_{T}$ vs. the differential Stark shift $\delta_{clock}(\theta=0)$. Different data points were measured at different trap powers. (a) $\textbf{B}$ is rotated between $\hat{x}$ and $\hat{z}$.
(b) $\textbf{B}$ is rotated between $\hat{y}$ and $\hat{z}$.
The mean values measured for each trap power are shown by the filled blue circles. The error bars are given by $\sigma/\sqrt{N}$ where N is the number of measurements at each trap power and $\sigma$ is their standard deviation. The uncertainties on the total shift, plotted on the figure abscissa, are similar in magnitude but cannot be seen on this scale. The blue solid line is a maximum likelihood linear fit to the data which yields the values given in Eq. 7. The grey-shaded area contains all the linear relations within $\chi^{2}=1$.}
\label{fig:figure3}
\end{figure*}

Note that since $\textbf{E}\perp\textbf{B}$ for both $\textbf{B}\parallel\hat{x},\hat{y}$, we obtain two equivalent data sets for $\delta_T$ vs. trap power (or equivalently $\delta_{clock}(\theta=0)$) shown in Fig.\ref{fig:figure3}.
We fit each data set with a linear relation $\delta_T=m\delta_{clock}(\theta=0)+b$, where $b$ accounts for possible systematic shifts.

Taking into account the statistical spread of the data in both $\delta_{clock}(\theta=0)$ and $\delta_T$, we get
\begin{eqnarray}
&& m_{x}=0.0125(10); \ b_{x}= 0.02(14)\\ \nonumber
&& m_{y}=0.0139(9); \ b_{y}=-0.21(13), \label{eq:mvalues}
\end{eqnarray}
with the uncertainty being one standard deviation to either side.
The subscripts $x$ and $y$ refer to the two different data sets of $\delta_{T}$.
The two estimated slopes disagree on order of one statistical uncertainty.
Having no preference to either one of the two configurations, we treat both equally and take the average slope as our result.
We take a conservative approach and combine the confidence intervals of the two measurements, thus getting
\begin{eqnarray*}
m=0.0132_{-0.0017}^{+0.0015} & ; & b=-0.09_{-0.24}^{+0.26}
\end{eqnarray*}

Several systematic effects are discussed below.
First, the clock transition's second order Zeeman shift $\delta_B$.
During the measurement we repeatedly measured and calibrated the magnetic field before taking each set of three Ramsey traces.
The calibration itself was done by measuring the shift of the first order magnetic sensitive transition and adjusting the current in the compensation coils such that it was $50\pm2\,\text{kHz}$.
The second order Zeeman shift, assumed common to all measurements was thus $\delta_{B}=2.93\pm0.23$ Hz.
Any constant error in the measured values of $B_x$, $B_y$ or $B_z$ will only serve to shift the linear relation of Fig.\ref{fig:figure3}, and will therefore only effect the bias, $b$, with no systematic shift in $m$.

Some systematic effects could result from deviations from the idealized model of harmonically trapped non-interacting atoms. Interactions between the atoms will result in density shifts.
We have minimized the uncertainty in this shift by reducing the number of trapped atoms to a minimum while maintaining a reasonable signal.
Furthermore, the resulting low collision rate improved the approximation of non-interacting atoms.
The density shifts in our measurement were always below $-0.5$\,Hz \cite{FertigGibble_PRL2000,HarberCornell_PRA2002} and the uncertainty in this shift is well below $0.1$\,Hz.
A source of error could arise if the number of trapped atoms is systematically over or under-estimated,
as the number of trapped atoms monotonically increased with the trap depth.
Nevertheless, even for densities an order of magnitude larger, and linearly dependent on the trap power, $m$ would have increased by less than 1\%.
We therefore conservatively estimate  1\% uncertainty due to this possible systematic effect.

To estimate the effect of trap inharmonicity we compare a more realistic Gaussian potential to the idealized harmonic trap.
At a distance from the trap bottom that equals to the extent of the atomic cloud, the leading order difference between the Gaussian and harmonic potentials relative to the total potential is $\frac{\Delta U}{U}\approx\frac{1}{8}(\frac{k_BT}{U_0})^2$.
This also limits the fractional change in the measured $\delta_{clock}$ due to trap inharmonicity, and is estimated to be smaller than 0.5\%.
Since this small inharmonicity mainly effects the broadening of the observed lineshape rather than its overall shift, and moreover, does not depend on the direction of the magnetic field, we neglect its effect on our result.

A third systematic error is due to possible misalignment of the magnetic coils with respect to $\hat{x}$, $\hat{y}$ and $\hat{z}$.
A small alignment error $\delta\theta\ll1$ will reduce the slope as  $m_{misaligned}\approx m_{aligned}(1-\delta\theta^{2})$.
We estimate the angular alignment error between the "x" magnetic bias coil and $\hat{x}$ to be $\delta\theta\lesssim 50$\,mrad.
The orthogonality of the three magnetic bias coils was verified by first calibrating each of them to a magnetic field corresponding to a first order Zeeman shift of 40kHz and then activating them in pairs.
The magnetic field from each perpendicular pair is then $B_{pair}=\sqrt{2}B_{coil}=56.6\,\text{kHz}$.
The measured fields are within 1kHz of this value, and thus the orthogonality between the coils is estimated to be better than 50\,mrad.
Combining the estimates for the alignment and orthogonality errors we bound this systematic effect by 0.5\%.

Another systematic decrease of the measured $m$ arises from a spread in the laser electric field direction across the atomic cloud.
The atoms are spread in a trap formed by a tightly focused beam, and if the cloud is too large, it may sample regions far from the focus, where the wavefront is no longer flat.
In our experiment the angle between different polarization vectors across the cloud can be bounded by
$\delta\theta^{2}<2.25\times10^{-4}$, and therefore negligible.

Another systematic shift is due to potential ellipticity in the trap laser polarization.
With an imperfect linear polarization, $\textbf{E}$ is never fully parallel or perpendicular to $\textbf{B}$, thereby lowering the measured value of $\delta_T$.
We measured the polarization linearity of the laser to be better than 1:50, bounding the effect by 4\%.

Finally, since our light is quasi-static and its polarization is, to a good approximation, linear any systematic shift due to a vector polarizability is negligible as well \cite{Derevianko_Magic2010PRA}.

After accounting for all the systematic shifts and using Eq.\ref{eq:measurable_s_t_ratio} we evaluate
\begin{equation*}
\frac{\alpha_{12}}{\alpha_{10}}=-0.0177\,_{-0.002}^{+0.0023}{}_{stat}\,_{-0.0011}^{+0.000}{}_{syst}.
\end{equation*}
Using a previous result for the value of $\alpha_{10}=0.930(3)\ 10^{-10}\text{Hz}/(\text{V}/\text{m})^2$ \cite{Safronova_PRA2010}, $\alpha_{12}$ is calculated to be
\begin{equation*}
\alpha_{12}=(-1.65\,_{-0.19}^{+0.21}{}_{stat}\,_{-0.11}^{+0.00}{}_{syst})\times 10^{-12}\text{Hz}/(\text{V}/\text{m})^2.
\end{equation*}

The spin-dipolar contribution to the tensor differential polarizability was calculated in \cite{DzubaFlambaumDerevianko_PRA2010} to be $\alpha_{12} = -1.48\times 10^{-12}\text{Hz}/(\text{V}/\text{m})^2$, consistent with our measurement. A previous measurement of the tensor polarizability in Rb$^{87}$ measured the shift of the flop-in transition under a static electric field which depends on $\alpha_{12}+\alpha_{02}$ \cite{GouldWeisskopf_PR1969}. Combining both measurements we extract the contribution of the interaction with the nuclear electric-quadrupole moment, $\alpha_{02} = -0.26(24)\times 10^{-12}\text{Hz}/(\text{V}/\text{m})^2$, again consistent with the calculated value of $\alpha_{02} = -0.34\times 10^{-12}\text{Hz}/(\text{V}/\text{m})^2$ \cite{DzubaFlambaumDerevianko_PRA2010}.

In conclusion, we have isolated and measured the spin-dipolar contribution to the tensor polarizability of Rb$^{87}$  ground state. Only the spin-dipolar contribution is relevant when considering shifts to the microwave clock transition and therefore it is of practical importance to atomic clocks.  By comparing our result to other tensor shift measurements we were able to determine the contribution of the electric-quadrupole interaction as well. Our results confirm recent calculations that relied on advanced quantum many-body methods.

We thank Nir Davidson and Ido Almog for many helpful discussions.
This work was financially supported by the US-Israel Binational Science Foundation, the Crown Photonics Center, ICore - Israeli excellence center "circle of light" and the European Research Council (consolidator grant 616919-Ionology).


\begin{thebibliography}{14}
\expandafter\ifx\csname natexlab\endcsname\relax\def\natexlab#1{#1}\fi
\expandafter\ifx\csname bibnamefont\endcsname\relax
  \def\bibnamefont#1{#1}\fi
\expandafter\ifx\csname bibfnamefont\endcsname\relax
  \def\bibfnamefont#1{#1}\fi
\expandafter\ifx\csname citenamefont\endcsname\relax
  \def\citenamefont#1{#1}\fi
\expandafter\ifx\csname url\endcsname\relax
  \def\url#1{\texttt{#1}}\fi
\expandafter\ifx\csname urlprefix\endcsname\relax\def\urlprefix{URL }\fi
\providecommand{\bibinfo}[2]{#2}
\providecommand{\eprint}[2][]{\url{#2}}

\bibitem[{\citenamefont{Dzuba et~al.}(2010)\citenamefont{Dzuba, Flambaum,
  Beloy, and Derevianko}}]{DzubaFlambaumDerevianko_PRA2010}
\bibinfo{author}{\bibfnamefont{V.~A.} \bibnamefont{Dzuba}},
  \bibinfo{author}{\bibfnamefont{V.~V.} \bibnamefont{Flambaum}},
  \bibinfo{author}{\bibfnamefont{K.}~\bibnamefont{Beloy}}, \bibnamefont{and}
  \bibinfo{author}{\bibfnamefont{A.}~\bibnamefont{Derevianko}},
  \bibinfo{journal}{Phys. Rev. A} \textbf{\bibinfo{volume}{82}},
  \bibinfo{pages}{062513} (\bibinfo{year}{2010}).

\bibitem[{\citenamefont{Rosenbad and Itano}(2006)}]{Rosenband2006}
\bibinfo{author}{\bibfnamefont{T.}\bibnamefont{Rosenband}}, \bibinfo{author}{\bibfnamefont{W.~M.}\bibnamefont{Itano}}, \bibinfo{author}{\bibfnamefont{P.~O.}\bibnamefont{Schmidt}}, \bibinfo{author}{\bibfnamefont{D.~B.}\bibnamefont{Hume}}, \bibinfo{author}{\bibfnamefont{J.~D.~J.}\bibnamefont{Koelemeij}}, \bibinfo{author}{\bibfnamefont{J.~C.}\bibnamefont{Bergquist}}, \bibnamefont{and}
\bibinfo{author}{\bibfnamefont{D.~J.} \bibnamefont{Wineland}},
\bibinfo{journal}{ Proc. 20th European Time and
Frequency Forum},
  \bibinfo{pages}{289} (\bibinfo{year}{2006}).

\bibitem[{\citenamefont{Sherman and Fortson}(2008)}]{Sherman2008}
\bibinfo{author}{\bibfnamefont{J.~A.}\bibnamefont{Sherman}}, \bibinfo{author}{\bibfnamefont{A.}\bibnamefont{Andalkar}}, \bibinfo{author}{\bibfnamefont{W.}\bibnamefont{Nagourney}}, \bibnamefont{and}
\bibinfo{author}{\bibfnamefont{E.~N.} \bibnamefont{Fortson}},
\bibinfo{journal}{Phys. Rev. A} \textbf{\bibinfo{volume}{78}},
\bibinfo{pages}{052514} (\bibinfo{year}{2008}).

\bibitem[{\citenamefont{Porsev and Ye}(2008)}]{Porsev2008}
\bibinfo{author}{\bibfnamefont{S.~G.}\bibnamefont{Porsev}}, \bibinfo{author}{\bibfnamefont{A.~D.}\bibnamefont{Ludlow}}, \bibinfo{author}{\bibfnamefont{M.~M.}\bibnamefont{Boyd}}, \bibnamefont{and}
\bibinfo{author}{\bibfnamefont{J.} \bibnamefont{Ye}},
\bibinfo{journal}{Phys. Rev. A} \textbf{\bibinfo{volume}{78}},
\bibinfo{pages}{032508} (\bibinfo{year}{2008}).

\bibitem[{\citenamefont{Holmgren and Cronin}(2012)}]{Holmgren2012}
\bibinfo{author}{\bibfnamefont{W.~F.}\bibnamefont{Holmgren}}, \bibinfo{author}{\bibfnamefont{R.}\bibnamefont{Trubko}}, \bibinfo{author}{\bibfnamefont{I.}\bibnamefont{Hromada}}, \bibnamefont{and}
\bibinfo{author}{\bibfnamefont{A.~D.} \bibnamefont{Cronin}},
\bibinfo{journal}{Phys. Rev. Lett.} \textbf{\bibinfo{volume}{109}},
\bibinfo{pages}{243004} (\bibinfo{year}{2012}).

\bibitem[{\citenamefont{Angel and Sandars}(1968)}]{SandarsAngel_RSPA1968}
\bibinfo{author}{\bibfnamefont{J.~R.~P.} \bibnamefont{Angel}} \bibnamefont{and}
  \bibinfo{author}{\bibfnamefont{P.~G.~H.} \bibnamefont{Sandars}},
  \bibinfo{journal}{Proceedings of the Royal Society of London. Series A.
  Mathematical and Physical Sciences} \textbf{\bibinfo{volume}{305}},
  \bibinfo{pages}{125} (\bibinfo{year}{1968}).

\bibitem[{\citenamefont{Safronova et~al.}(2010)\citenamefont{Safronova, Jiang,
  and Safronova}}]{Safronova_PRA2010}
\bibinfo{author}{\bibfnamefont{M.~S.} \bibnamefont{Safronova}},
  \bibinfo{author}{\bibfnamefont{D.}~\bibnamefont{Jiang}}, \bibnamefont{and}
  \bibinfo{author}{\bibfnamefont{U.~I.} \bibnamefont{Safronova}},
  \bibinfo{journal}{Phys. Rev. A} \textbf{\bibinfo{volume}{82}},
  \bibinfo{pages}{022510} (\bibinfo{year}{2010}).

\bibitem[{\citenamefont{Guena et~al.}(2010)\citenamefont{Guena, Rosenbusch,
  Laurent, Abgrall, Rovera, Santarelli, Tobar, Bize, and
  Clairon}}]{GuenaClarionRosenbusch_IEEE2010}
\bibinfo{author}{\bibfnamefont{J.}~\bibnamefont{Guena}},
  \bibinfo{author}{\bibfnamefont{P.}~\bibnamefont{Rosenbusch}},
  \bibinfo{author}{\bibfnamefont{P.}~\bibnamefont{Laurent}},
  \bibinfo{author}{\bibfnamefont{M.}~\bibnamefont{Abgrall}},
  \bibinfo{author}{\bibfnamefont{D.}~\bibnamefont{Rovera}},
  \bibinfo{author}{\bibfnamefont{G.}~\bibnamefont{Santarelli}},
  \bibinfo{author}{\bibfnamefont{M.}~\bibnamefont{Tobar}},
  \bibinfo{author}{\bibfnamefont{S.}~\bibnamefont{Bize}}, \bibnamefont{and}
  \bibinfo{author}{\bibfnamefont{C.}~\bibnamefont{Clairon}},
  \bibinfo{journal}{Ultrasonics, Ferroelectrics and Frequency Control, IEEE
  Transactions on} \textbf{\bibinfo{volume}{57}}, \bibinfo{pages}{647}
  (\bibinfo{year}{2010}), ISSN \bibinfo{issn}{0885-3010}.

\bibitem[{\citenamefont{Mowat}(1972)}]{Mowat_PRA1972}
\bibinfo{author}{\bibfnamefont{J.~R.} \bibnamefont{Mowat}},
  \bibinfo{journal}{Phys. Rev. A} \textbf{\bibinfo{volume}{5}},
  \bibinfo{pages}{1059} (\bibinfo{year}{1972}).

\bibitem[{\citenamefont{Simon et~al.}(1998)\citenamefont{Simon, Laurent, and
  Clairon}}]{SimonClarion_PRA1998}
\bibinfo{author}{\bibfnamefont{E.}~\bibnamefont{Simon}},
  \bibinfo{author}{\bibfnamefont{P.}~\bibnamefont{Laurent}}, \bibnamefont{and}
  \bibinfo{author}{\bibfnamefont{A.}~\bibnamefont{Clairon}},
  \bibinfo{journal}{Phys. Rev. A} \textbf{\bibinfo{volume}{57}},
  \bibinfo{pages}{436} (\bibinfo{year}{1998}).

\bibitem[{\citenamefont{Ulzega thesis}(2006)\citenamefont{Ulzega}}]{UlzegaThesis}
\bibinfo{author}{\bibfnamefont{S.}~\bibnamefont{Ulzega}},
  \bibinfo{journal}{PhD Thesis, University of Fribourg, Switzerland}
 (\bibinfo{year}{2006}).

\bibitem[{\citenamefont{Ulzega et~al.}(2006)\citenamefont{Ulzega, Hofer,
  Moroshkin, and Weis}}]{UlzegaWeis2006}
\bibinfo{author}{\bibfnamefont{S.}~\bibnamefont{Ulzega}},
  \bibinfo{author}{\bibfnamefont{A.}~\bibnamefont{Hofer}},
  \bibinfo{author}{\bibfnamefont{P.}~\bibnamefont{Moroshkin}},
  \bibnamefont{and} \bibinfo{author}{\bibfnamefont{A.}~\bibnamefont{Weis}},
  \bibinfo{journal}{EPL (Europhysics Letters)} \textbf{\bibinfo{volume}{76}},
  \bibinfo{pages}{1074} (\bibinfo{year}{2006}).

\bibitem[{\citenamefont{Gould et~al.}(1969)\citenamefont{Gould, Lipworth, and
  Weisskopf}}]{GouldWeisskopf_PR1969}
\bibinfo{author}{\bibfnamefont{H.}~\bibnamefont{Gould}},
  \bibinfo{author}{\bibfnamefont{E.}~\bibnamefont{Lipworth}}, \bibnamefont{and}
  \bibinfo{author}{\bibfnamefont{M.~C.} \bibnamefont{Weisskopf}},
  \bibinfo{journal}{Phys. Rev.} \textbf{\bibinfo{volume}{188}},
  \bibinfo{pages}{24} (\bibinfo{year}{1969}).

\bibitem[{\citenamefont{Ospelkaus et~al.}(2003)\citenamefont{Ospelkaus,
  Rasbach, and Weis}}]{OpelklausWeis2003}
\bibinfo{author}{\bibfnamefont{C.}~\bibnamefont{Ospelkaus}},
  \bibinfo{author}{\bibfnamefont{U.}~\bibnamefont{Rasbach}}, \bibnamefont{and}
  \bibinfo{author}{\bibfnamefont{A.}~\bibnamefont{Weis}},
  \bibinfo{journal}{Phys. Rev. A} \textbf{\bibinfo{volume}{67}},
  \bibinfo{pages}{011402} (\bibinfo{year}{2003}).

\bibitem[{\citenamefont{Takekoshi et~al.}(1995)\citenamefont{Takekoshi, Yeh,
  and Knize}}]{TakekoshiKnize_OPtCom1995}
\bibinfo{author}{\bibfnamefont{T.}~\bibnamefont{Takekoshi}},
  \bibinfo{author}{\bibfnamefont{J.~R.} \bibnamefont{Yeh}}, \bibnamefont{and}
  \bibinfo{author}{\bibfnamefont{R.~J.} \bibnamefont{Knize}},
  \bibinfo{journal}{Optics Communications} \textbf{\bibinfo{volume}{114}},
  \bibinfo{pages}{421 } (\bibinfo{year}{1995}), ISSN \bibinfo{issn}{0030-4018}.

\bibitem[{\citenamefont{Kuhr et~al.}(2005)\citenamefont{Kuhr, Alt, Schrader,
  Dotsenko, Miroshnychenko, Rauschenbeutel, and
  Meschede}}]{KuhrMeschede_PRA2005}
\bibinfo{author}{\bibfnamefont{S.}~\bibnamefont{Kuhr}},
  \bibinfo{author}{\bibfnamefont{W.}~\bibnamefont{Alt}},
  \bibinfo{author}{\bibfnamefont{D.}~\bibnamefont{Schrader}},
  \bibinfo{author}{\bibfnamefont{I.}~\bibnamefont{Dotsenko}},
  \bibinfo{author}{\bibfnamefont{Y.}~\bibnamefont{Miroshnychenko}},
  \bibinfo{author}{\bibfnamefont{A.}~\bibnamefont{Rauschenbeutel}},
  \bibnamefont{and} \bibinfo{author}{\bibfnamefont{D.}~\bibnamefont{Meschede}},
  \bibinfo{journal}{Phys. Rev. A} \textbf{\bibinfo{volume}{72}},
  \bibinfo{pages}{023406} (\bibinfo{year}{2005}).

\bibitem[{\citenamefont{Fertig and Gibble}(2000)}]{FertigGibble_PRL2000}
\bibinfo{author}{\bibfnamefont{C.}~\bibnamefont{Fertig}} \bibnamefont{and}
  \bibinfo{author}{\bibfnamefont{K.}~\bibnamefont{Gibble}},
  \bibinfo{journal}{Phys. Rev. Lett.} \textbf{\bibinfo{volume}{85}},
  \bibinfo{pages}{1622} (\bibinfo{year}{2000}).

\bibitem[{\citenamefont{Harber et~al.}(2002)\citenamefont{Harber, Lewandowski,
  McGuirk, and Cornell}}]{HarberCornell_PRA2002}
\bibinfo{author}{\bibfnamefont{D.~M.} \bibnamefont{Harber}},
  \bibinfo{author}{\bibfnamefont{H.~J.} \bibnamefont{Lewandowski}},
  \bibinfo{author}{\bibfnamefont{J.~M.} \bibnamefont{McGuirk}},
  \bibnamefont{and} \bibinfo{author}{\bibfnamefont{E.~A.}
  \bibnamefont{Cornell}}, \bibinfo{journal}{Phys. Rev. A}
  \textbf{\bibinfo{volume}{66}}, \bibinfo{pages}{053616}
  (\bibinfo{year}{2002}).

\bibitem[{\citenamefont{Derevianko}(2010)}]{Derevianko_Magic2010PRA}
\bibinfo{author}{\bibfnamefont{A.}~\bibnamefont{Derevianko}},
  \bibinfo{journal}{Phys. Rev. A} \textbf{\bibinfo{volume}{81}},
  \bibinfo{pages}{051606} (\bibinfo{year}{2010}).


\end{thebibliography}
\end{document}